\documentclass[aps,twocolumn,prl,showpacs]{revtex4}

\usepackage{graphicx}
\def\bm#1{{\mbox{\boldmath $#1$}}}
\def\omegaBP{\omega_{\mathrm{BP}}}
\def\rhoc{\rho_{\mathrm c}}

\begin{document}

\title{Boson peak in an harmonic scalar model}

\author{T.~S.~Grigera}
\author{V.~Mart\'\i{}n-Mayor} 
\author{G.~Parisi}
\affiliation{Dipartimento di Fisica, Universit\`a di Roma ``La
Sapienza'', P.le Aldo Moro 2, 00185 Roma, Italy INFN sezione di
Roma - SMC and UdR1 of INFM}
\author{P.~Verrocchio}
\affiliation{Dipartimento di Fisica, Universit\`a
di Trento, Via Sommarive, 14, 38050 Povo, Trento, Italy INFM unit\`a
di Trento}
\date{\today}
\begin{abstract}
We study, analytically and numerically, an off-lattice
model of scalar harmonic vibrations for structural glasses.  The
model has a Boson-Peak which we argue can be considered as a prototype
for materials with a Boson Peak frequency that decreases with lowering
temperature. The density evolution of the Boson Peak in silica is
qualitatively reproduced. The dispersion relation is linear at the
Boson Peak frequency, in agreement with experiments.  In our model the
Boson Peak is a precursor of a nearby mechanical instability. The
Boson Peak is built up by the hybridization of the sound waves with
extended, but non propagating (in the sense of a flat dispersion
relation), modes.
\end{abstract}
\pacs{PACS 61.43.Fs, 63.50.+x}
\maketitle

The vibrational dynamics of glasses at the frequencies that control
the specific heat in the $1\,$K--$100\,$K range is a long-standing
physical problem. At the lowest temperatures, an anomalous linear
dependence on $T$ can be reproduced by the two-level
model~\cite{twolevel}. At slightly higher temperatures, another
deviation from the Debye $T^3$ law is found: the specific heat divided
by $T^3$ shows a peak near $10\,$K \cite{pohl81}. This peak points to
the existence of an excess of vibrational states at frequencies
$\omega \sim 1\,$THz, which shows as a peak in the plot of
$g(\omega)/\omega^2$ ($g(\omega)$ being the vibrational density of
states, as obtained by Raman or inelastic neutron scattering). Since
the scattering intensity at the peak scales in temperature with Bose
statistics, the peak has become known as the {\em Boson peak}
(BP). The physical origin of the BP is a matter of a lively debate
within the experimental and theoretical glass community.

The recent experimental efforts to elucidate the nature of the high
frequency excitations in glasses have produced a wealth of inelastic
scattering data (X-ray, Raman and neutrons). These experiments have
found a BP on almost all studied glassformers, although of an
intensity that can vary substantially among different materials.  Two
features of the BP should be remarked. First, the peak frequency
$\omegaBP$ is several times smaller than any natural frequency scale,
like the Debye frequency or the band edge. Moreover, where good data
for the dispersion relation $\omega(p)$ (determined from the the
Brillouin peak in the dynamic structure factor, which probes only
longitudinal modes) are available, it has been checked that
$\omega(p)$ is still a linear function of the momentum $p$ at
$\omegaBP$. This is the case for glycerol, LiCl:6H$_2$O
\cite{glycerol}, Ca$_{0.4}$K$_{0.6}$(NO$_3$)$_{1.4}$, B$_2$O$_3$
\cite{Matic01}, silica \cite{silica} and polybutadiene
\cite{Fioretto99} (the interpretation of scattering experiments in
silica has been controversial \cite{SilicaDebate}).  Second, although
less experimental data are available, it seems to be the rule that
$\omegaBP$ shifts to lower frequencies on heating, while the BP
intensity grows. This is the case in polybutadiene, polystyrene
\cite{Sokolov95}, LiCl \cite{Tao91} and B$_2$O$_3$
\cite{Engberg99}. Silica, however, shows the opposite behavior
\cite{Sokolov95,Wischnewski98}.  Yet, in silica the BP evolution upon
increasing the density has been studied experimentally
\cite{SILICADENSITYEXP} and in simulations \cite{SILICADENSITYNUM}.
In close agreement with the results presented here, it was found that
$\omegaBP$ shifts to larger frequencies and the BP looses intensity
when the density grows.  The BP has also been identified in the
dynamic structure factor at high exchanged momentum, both
experimentally \cite{talks} and in simulations \cite{Horbach01}, which
reflects the fact that the dynamic structure factor
tends to the density of state (DOS) at high momentum
\cite{Martin-Mayor01,Grigera01}.

Perhaps the only widely-agreed upon statement about the BP is that it,
as well as the whole dynamic structure factor in this frequency
region, can be understood invoking only harmonic vibrations
\cite{Horbach99,Taraskin99,Ruocco00,Gotze00,Grigera01,Schirmacher98,
vanHove,Kantelhardt01}
(see however the work on soft-potentials \cite{soft-potentials} for a
dissenting view).  Otherwise, the BP has been variously interpreted as
arising from mixing of longitudinal and transversal modes
\cite{Horbach01,Sampoli97,Ruocco00,Matic01}, hybridization of optical
and acoustic modes \cite{Taraskin99}, a combination of the
level repulsion due to disorder and van Hove singularities
\cite{vanHove}, an associated mechanical instability
\cite{Grigera01,Gotze00} (as suggested by the temperature dependence
of $\omegaBP$), the presence of a Ioffe-Regel crossover at $\omegaBP$
\cite{Foret96,Parshin01}, or from the scattering of sound-waves with
localized anharmonic
vibrations\cite{soft-potentials,Wischnewski98}. The BP has also been
identified with the anomalous oscilation peak found in Mode Coupling
Theory as adapted to the non-ergodic phase \cite{Gotze00,Theenhaus01}.

In this Letter we discuss analitically and numerically an off-lattice
model of {\em scalar} harmonic vibrations, which is a problem on
random-matrix theory~\cite{Metha91}, showing that it has a Boson peak
with the above-mentioned experimentally identified
characteristics. Thus it is unnecessary to invoke anharmonicity or
transverse modes to explain the BP (although they are certainly
present in real glasses), which instead, as we will show, arises from
the hybridization of {\em extended} but non-propagating modes with
sound waves, induced by a mechanical instability.  Our model
\cite{Mezard99,Martin-Mayor01,Grigera01} consists in particles
oscillating harmonically around (disordered) equilibrium positions,
along a {\em fixed direction} {\boldmath $u$} (i.e. we neglect all
transverse modes): $\bm{x}_i(t) = \bm{x}^{\mathrm{eq}} + \bm{u}
\varphi_i(t),$ $1,\ldots,N$. The potential energy is $V = {1\over2}
\sum_{i \ne j}^N f(\bm{x}_i^{\mathrm{eq}} - \bm{x}_j^{\mathrm{eq}})
(\varphi_i - \varphi_j)^2 = \sum_{i,j}^N \varphi_i M_{ij} \varphi_j$,
where $f$ is the second derivative of the pair potential and the
dynamical matrix matrix (Hessian) is
\begin{equation}
M_{ij}=\delta_{ij} \sum_{k=1}^N
f({\mbox{\boldmath$x$}}_i^{\mathrm{eq}}-{\mbox{\boldmath$x$}}_k^{\mathrm{eq}})
-
f({\mbox{\boldmath$x$}}_i^{\mathrm{eq}}-{\mbox{\boldmath$x$}}_j^{\mathrm{eq}}).
\label{HESSIANO}
\end{equation}
The model is completely defined when $f$ and the distribution of the
$\bm{x}^{\mathrm{eq}}$ are chosen. We take the latter uniform in the
whole volume, while $f$ is taken to be regular at short distances. As
discussed in Ref.~\cite{Martin-Mayor01}, this is a good
first approximation to the correlations of the $\bm{x}^{\mathrm{eq}}$.
We shall consider the family of functions
\begin{equation}
f_\alpha (\bm{r})=(1-\alpha \bm{r}^2/\sigma^2)
e^{-\bm{r}^2/(2\sigma^2)},
\end{equation}
where $0\le \alpha \le 0.2$ (the upper bound has to be imposed in
order to guarantee a positive sound velocity). When $\alpha=0$
(Gaussian case), the hessian is strictly positive. When $\alpha>0$, we
have a stable elastic solid at high particle-number density, $\rho$,
while at low enough $\rho$, typical interparticle distances will be
large enough to allow negative eigenvalues (imaginary
frequencies). Therefore the density controls the appearence of a
mechanical instability in this model: there is a phase-transition from
a region where the short time dynamics is ruled by a positive hessian
(high $\rho$) to the one where an extensive number of negative
eigenvalues is found (low $\rho$). The counterpart in real glasses of
that transition is the mode-coupling transition~\cite{TDYN} which is
rather a crossover, but it has been recently argued~\cite{saddles}
to correpond to a well defined phase-transition on the topological
properties of the potential energy landscape. It must also be said
that in disordered lattices models~\cite{vanHove,Schirmacher98}, an
instability analogous to ours is found if the disorder is strong
enough. Nevertheless, the bump in $g(\omega)/\omega^2$ found on the
lattice is intrinsically tied to a characteristic frequency (the
van-Hove singularity, see~\cite{vanHove,Grigera01} and below).

\begin{figure}
\includegraphics[angle=90,width=\columnwidth]{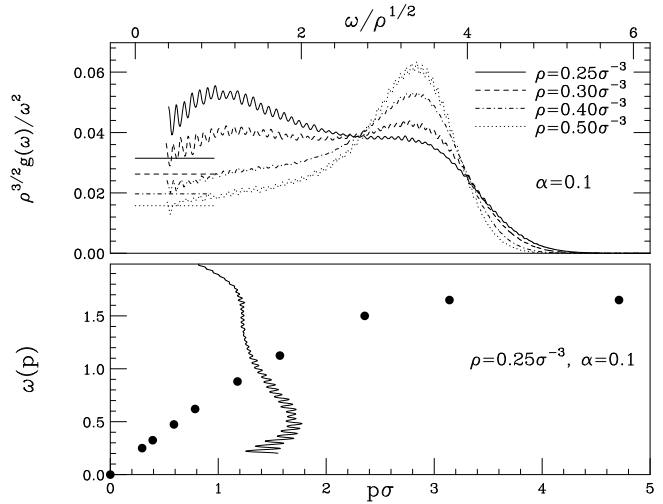}
\caption{Top: BP raising in our model
with decreasing $\rho$; the horizontal lines indicate the Debye
value.  Bottom: numerical dispersion relation from the dynamic
structure factor ($g(\omega)/\omega^2$ rotated 90
degrees is also ploted).  $\omega(p)$ is linear at $\omegaBP$ and gets
flat at the high-frequency bump in $g(\omega)/\omega^2$.}
\label{NO-GAUSSIAN-NUMERICO}
\end{figure}

We first compute the DOS of our model numerically for $\alpha=0.1$,
using the method of moments \cite{Benoit92} with a box of side
$L=128\,\sigma$ (more than $5\times 10^5$ particles), which allows to
reconstruct the spectrum up to very low frequencies. The critical
$\rho$ is difficult to locate, but at $\rho=0.05\sigma^{-3}$ imaginary
frequencies are clearly found.  In Fig.~\ref{NO-GAUSSIAN-NUMERICO}
(top) we show the DOS divided by $\omega^2$ at several densities. We
have supressed the smallest frequencies, which are unreliable due to
finite-size effects, and used the reduced form \cite{Martin-Mayor01}
$\rho^{3/2} g(\omega)/\omega^2$ vs.\ $\omega/\sqrt{\rho}$ in order to
compare different densities on the same plot.  A peak in
$g(\omega)/\omega^2$ arises on approaching the instability (i.e. on
lowering the density). As $\rho$ is reduced, the peak grows (relative
to the Debye value, also plotted) and moves to lower $\omega$. The
peak lies always in the linear region of the dispersion
relation (Fig.~\ref{NO-GAUSSIAN-NUMERICO}, bottom). Since decreasing
$\rho$ plays the role of increasing temperature in our model, 
the experimental features of the BP are reproduced, without
transverse or optical modes or anharmonicity, as claimed.

\begin{figure}[h!]
\includegraphics[angle=90,width=\columnwidth]{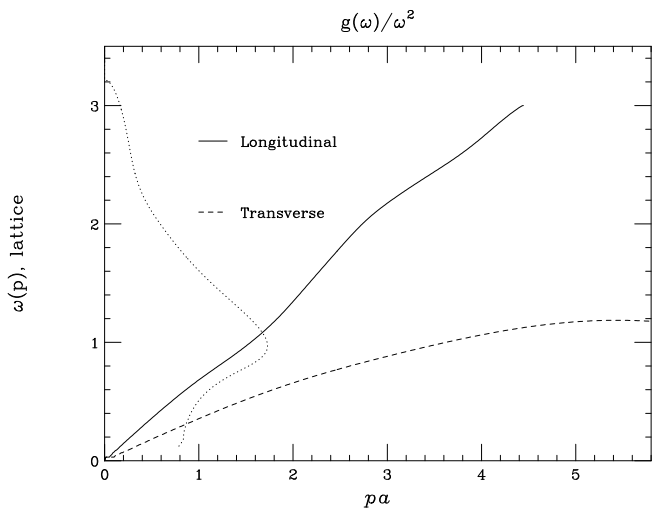}
\caption{Dispersion relation of a heavily disordered ($\Delta=1.0$)
FCC lattice model~\protect\cite{vanHove} along the $(1,1,1)$ direction,
and (rotated) $g(\omega)/\omega^2$, from the CPA approximation ($a$ is
the lattice spacing). The BP is at the end of the transverse
branch.}
\label{SONIDO-ANALITICO}
\end{figure}

Let us discuss analytically the BP found in our
model. Within our random-matrix approach
\cite{Mezard99,Martin-Mayor01}, the DOS is obtained from the resolvent
$G(p,z) \equiv (1/N) \sum_{jk} \overline{ \exp[\mathrm{i}
\mbox{\boldmath $p$} \cdot ( {\mbox{\boldmath$x$}}^{\mathrm eq}_j-
{\mbox{\boldmath$x$}}^{\mathrm eq}_k ) ] [(z-M)^{-1}]_{jk} }$, where the
overline stands for the average over $\bm{x}^{\mathrm{eq}}$. Defining
${\cal G}(z) = G(p=\infty,z)$, the DOS is
$g(\omega) = - {2\omega \over \pi} \, \mathrm{Im}\, {\cal G} (\omega^2
+ i 0^+)$.
Expanding the resolvent in $1/\rho$ \cite{Martin-Mayor01} and
resumming an infinite subset of diagrams \cite{Grigera01}, a
non-linear equation is found for $G(p,z)$ that in the $p\to\infty$ limit
yields
\begin{equation}
\frac{1}{\rho{\cal G}(z)}=\frac{z}{\rho}-\hat f(0) - A {\cal G}(z) -\int
\!\!\! \frac{d^3q}{(2\pi)^3}\hat f^2(\mbox{\boldmath$q$})
G(\mbox{\boldmath$q$},z),
\label{DENSIDADDEESTADOS}
\end{equation} 
where $A= (2\pi)^{-3}\int \!\!\hat f^2(\mbox{\boldmath$q$})\,d^3q$ and
$\hat f(\bm{q})$ is the Fourier transform of $f(\bm{r})$. With this
equation, one needs to know the resolvent at all $q$ to obtain the
DOS, due to the last term in the r.h.s. This can be done by solving
numerically the self-consistent equation of
ref.~\cite{Grigera01}. Here we perform an approximate analysis, which
is more illuminating. The crudest approximation is to neglect this
term, in which case Eq.~\ref{DENSIDADDEESTADOS} is quadratic in ${\cal
G}$, and one easily finds a semicircular DOS, with center at
$\omega^2=\rho \hat f(0)$ and radius $2\sqrt{\rho A}$. Indeed, when
$\rho\to \infty$, the spectrum is made of plane waves, with dispersion
relation $\omega^2(p)= \rho(\hat f(0)-\hat
f(p))$,~\cite{Martin-Mayor01,Mezard99} (a continous elastic
medium). $\omega(p)$ saturates for large $p$ at $\omega^2=\rho\hat
f(0)$, yielding an enormous pile-up of extended (but non-propagating)
states which causes the DOS to be concentrated at this
frequency~\cite{Mezard99}. This is the glass analogue of a van-Hove
singularity.  At finite $\rho$, the density fluctuations of the
$\bm{x}^{\mathrm{eq}}$ split this degeneracy, and yield the
semicircular part of the spectrum at high frequency that can be
recognized in the upper panel of Fig.~\ref{NO-GAUSSIAN-NUMERICO}. But
the semicircular spectrum misses the Debye part, and a better
approximation is needed.  So we substitute $G$ in the last term of the
r.h.s. by the resolvent of the continuum elastic medium $G_0(z,p) = (
z - \omega^2(p))^{-1}$.  This is reasonable because the $f^2(q)$
factor makes low momenta dominate the integral, and due to
translational invariance $G(z,p) \approx G_0(z,p)$ in this
region~\cite{Martin-Mayor01}. We shall be looking at small $\omega$,
so to a good approximation
\begin{equation}
\int \!\! {d^3q \over (2\pi)^3} \, \hat f^2(q) G_0(q,z) \approx
- {1\over\rho} B - i {\rho \hat f^2(0) \over 4\pi c^3} \omega,
\end{equation}
where the sound velocity is $c=\sqrt{\rho \hat f''(0)/2}$, while $B$
is a positive constant. Then Eq.~(\ref{DENSIDADDEESTADOS}) is again
quadratic in ${\cal G}$, and can be solved to give
\begin{eqnarray}
&&{\cal G}(\omega^2+i0^+) \approx 
\frac{\omega^2 - \rho \hat f(0) + B + i \frac{\rho\hat f^2(0) \omega}{4\pi
c^3}} {2\rho A} \times \nonumber \\  
&&\times\left( 1 - \sqrt{ 1 - 
\frac{4\rho A}{[\omega^2 - \rho \hat f(0) + B + i \frac{\rho\hat f^2(0)
\omega}{4\pi c^3}]^2}} \right). \label{SMALLOMEGA}
\end{eqnarray}
We have two limiting cases. At high densities and low frequencies
($\rho \hat f(0) \gg \omega^2, B, 2\sqrt{\rho A}$), i.e.\ when the
semicircular part of the DOS does not reach low frequencies, the
square root can be Taylor-expanded, and one gets
\begin{equation}
g(\omega) \approx \frac{\omega^2}{2\pi \rho c^3}, \label{DEBYEPREFACTOR}
\end{equation}
which is precisely Debye's law. At small densities, on the other hand,
the center of the semicircle (which is at $\omega^2=\rho \hat f(0) -
B$) starts to be comparable to its radius ($\propto \sqrt{\rho}$),
implying that the states in the semicircle hybridize with the sound
waves. This is the source of the mechanical instability we find
numerically in Fig.~\ref{NO-GAUSSIAN-NUMERICO}. Mathematically, the
instability arises when ${\cal G}(0)$ develops an imaginary part. This
can only come from the square root in Eq.~(\ref{SMALLOMEGA}), and it
will happen for $\rho<\rhoc$, with $\rhoc$ fixed by the condition
$2 \sqrt{A\rhoc} + B = \rhoc \hat f(0).$
Now when $\rho \gtrsim \rhoc$ and $\omega \ll \omega^* = 2\pi
c^3\sqrt{\rhoc A}/(\rhoc \hat f^2(0))$, the square root in
Eq.~(\ref{SMALLOMEGA}) behaves as
\begin{equation}
\sqrt{ D(\rho-\rhoc) - i \omega/\omega^* },
\label{ORDEN}
\end{equation}
with $D$ a positive constant. One distinguishes two regimes:
\begin{itemize}
\item $\omega^* D (\rho-\rhoc) \ll \omega \ll \omega^*$: the
imaginary part of $\cal G$ is proportional to $\sqrt{\omega}$, 
the DOS being $g(\omega) \propto \omega^{3/2}$.
\item $\omega \ll \omega^* D (\rho-\rhoc) \ll \omega^*$:
the
imaginary part of $\cal G$ is proportional to 
$\omega$, and $g(\omega) =
\omega^2/(\omega^* \sqrt{\rhoc A D (\rho-\rhoc)})$. The DOS is
Debye-like with a prefactor arbitrarily larger than in
Eq.(\ref{DEBYEPREFACTOR}).
\end{itemize}
\begin{figure}
\includegraphics[angle=90,width=\columnwidth]{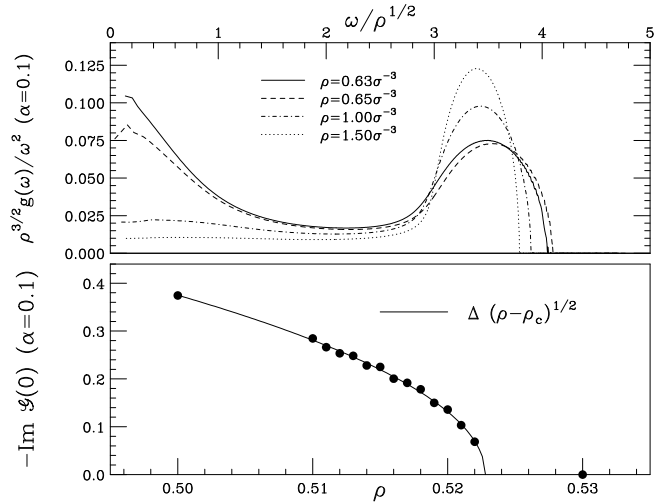}
\caption{Top: $g(\omega)/\omega^2$ from the numerical solution of the
self-consitent equation of \protect\cite{Grigera01}. Bottom: Order
parameter of the mechanical instability phase transition vs.\ density
(points). Solid line is a fit to the predicted behavior $\Delta
(\rhoc-\rho)^{1/2}$.}
\label{NO-GAUSSIAN-ANALITICO}
\end{figure}
One thus identify $\omegaBP$ with $\omega^* D (\rho-\rhoc)$ that,
on approaching the instability, becomes arbitrarily small compared to
any natural frequency scale.  The mechanical instability is a phase
transition, for which the order parameter is $-\mathrm{Im}\, {\cal
G}(0)$. From Eq.~(\ref{ORDEN}) we see that this order parameter
behaves as $(\rhoc-\rho)^\beta$, with $\beta=1/2$, like in mean-field
theories. In this sense, we can say that Eq.~(\ref{DENSIDADDEESTADOS})
is a mean-field theory for the model.  That our qualitative analysis
is correct can be checked looking at Fig.~\ref{NO-GAUSSIAN-ANALITICO},
where we show the DOS, and the behavior of $-\mathrm{Im}\, {\cal
G}(0)$ with $\rho$, both obtained from the self-consistent $G(q,z)$.
Note also that the analytical solution compares reasonably well with
the numerical solution (Fig.~\ref{NO-GAUSSIAN-NUMERICO}) on a
qualitative level (though it overestimates $\rhoc$, and likely finds
the wrong critical exponent).  Since the behavior of the propagator at
high momentum does not strongly affect the dispersion
relation~\cite{Grigera01}, we do not expect deviations from a linear
dispersion relation at $\omegaBP$.  This can be checked either
numerically (Fig.~\ref{NO-GAUSSIAN-NUMERICO}, bottom)
or solving numerically the self-consistency equations.  This is in
sharp contrast with lattice models for the
BP~\cite{vanHove,Schirmacher98}. Indeed, on the lattice the DOS is
given by the integral of the dynamic structure factor, which is
dominated by the Brillouin-peak. Thus the so-called lattice BP
reflects a non smooth behavior of the Brillouin-peak.  In fact, one
can check (Fig.~\ref{SONIDO-ANALITICO}) that in these models
the BP arise from a transverse van-Hove singularity that affects
strongly the longitudinal dispersion relation.

In summary, we have discussed an harmonic model for the high frequency
dynamics of glasses, without transverse modes. The model has a
mechanical instability transition controlled by the density.  We have
studied the model numerically, and analytically with a mean-field
theory for the instability transition. The vibrational spectrum
contains a BP which is the precursor of the transition.  The BP in our
model shares the main features of the experimental BP: it appears for
frequencies in the linear part of the dispersion relation and it
shifts towards arbitrarily low frequencies on approaching a mechanical
instability.  We also reproduces qualitatively the behaviour of the
silica BP when the density
changes~\cite{SILICADENSITYEXP,SILICADENSITYNUM} (the detailed theory
of the temperature evolution of the silica BP should consider
its negative thermal dilatation coefficient).  The BP is built from
the hybridization of sound waves with high frequency modes (extended
but non propagating) that get softer upon approaching the instability.
The analogous of our instability transition in nature is the
topological phase transition~\cite{saddles} that underlies the dynamic
crossover at the Mode Coupling temperature of real
glasses~\cite{TDYN}.  The precise nature of the high-frequency modes
that hybridize with the sound-waves is most likely material dependent
and non-universal: they could be
transverse~\cite{Horbach01,Sampoli97,Ruocco00,Matic01},
optical~\cite{Taraskin99}, or even longitudinal modes as in our
model. We believe however that the basic mechanism for the formation
of the BP uncovered in our model is common to most (if not all) the
structural glasses. Yet all real glasses do have transverse
excitations, and one could ask about generic new features introduced
by these modes.  We are currently working to extend the present theory
in this direction.

We thank G. Ruocco, O. Pilla and G. Viliani for discussions. TSG was
supported in part by CONICET (Argentina), and VMM by E.C. contract
HPMF-CT-2000-00450.

\end{document}